\documentclass[epjCONF]{svjour}
\usepackage{graphics}
\usepackage[varg]{txfonts} 
\usepackage[latin1]{inputenc}
\session-title{Conference title}
\begin{document}
\title{In-medium Properties of the $\eta^\prime$-Meson from Photonuclear Reactions}
\author{Mariana Nanova\inst{1} \fnmsep\thanks{\email{Mariana.Nanova@exp2.physik.uni-giessen.de}}\\ for the CBELSA/TAPS Collaboration}
\institute{II. Physikalisches Institut, University of Giessen, Germany}
\abstract{
Investigations of in-medium properties of the $\eta^\prime$-meson are presented. The photoproduction of $\eta^\prime$-mesons off ${}^{12}\textrm{C}, {}^{40}\textrm{Ca}, {}^{93}\textrm{Nb}$ and ${}^{208}\textrm{Pb}$ nuclei has been measured with the Crystal Barrel/TAPS detector system at the ELSA accelerator in Bonn. Comparing the meson yield per nucleon within a nucleus to the yield on a free nucleon, information on the absorption of the meson through inelastic channels in the nuclear medium and the related in-medium width of the meson is provided. From this so called transparency ratio we found the in-medium width of the $\eta^\prime$-meson to be of the order of 20 MeV.  The measured width is sufficently narrow to encourage a search for $\eta^\prime$ mesic states. Recent results on the momentum distribution of the transparency ratio and in-medium width of the $\eta^\prime$-meson are presented and discussed.\\ 
} 
\maketitle
\section{Introduction}
\label{intro}
	The origin of hadron masses is one of the open questions of QCD in the low-energy sector. The light pseudoscalar mesons ($\pi$, $K$, $\eta$) are the Nambu-Goldstone bosons which are associated with the spontaneous breaking of the QCD chiral symmetry. Introducing non zero masses of the current quarks, these mesons together with the heavier $\eta^\prime$(958) meson show a mass spectrum which is believed to be explained by the explicit flavor $SU(3)$ breaking and the axial $U_{A}(1)$ anomaly. There have been many theoretical studies discussing the idea that the chiral symmetry may be at least partially restored in a nuclear medium \cite{meissner,costa,jido} which might have an impact on the in-medium $\eta^\prime$ mass. Indirect evidence has been claimed for a dropping $\eta^\prime$ mass in the hot and dense matter formed in ultrarelativistic heavy-ion collisions at RHIC energies \cite{Csoergo}. \\
The  $\eta^\prime N$ scattering length has been estimated from the study of the $pp \to pp \eta^\prime$ cross section near threshold at COSY  \cite{Moskal,Moskal1}. A refined analysis of this reaction, comparing the cross section with that of the $pp \to pp \pi ^0$ reaction, concluded that the $\eta^\prime$ scattering length should be of the order of magnitude of that of the $\pi N$ interaction and hence $|a_{\eta^\prime N}|\sim 0.1$ fm \cite{Moskal1}. This indicates a rather weak $\eta^\prime N$ interaction. In \cite{oset_ramos} it was interpreted as a consequence of the particular dynamics of the $\eta^\prime$ as a singlet meson with a small octet admixture given by the $\eta$ and $\eta^\prime$ mixing angle.\\ 
An experimental approach to learn about the $\eta^\prime N$ interaction and the in-medium properties of the $\eta^\prime$ meson is the study of $\eta^\prime $ photoproduction off nuclei. The in-medium width of the $\eta^\prime $-meson can be extracted from the attenuation of the $\eta^\prime$-meson flux deduced from a measurement of the transparency ratio for a number of nuclei. Unless when removed by inelastic channels the $\eta^\prime$-meson will decay outside of the nucleus because of its long lifetime and thus its in-medium mass is not accessible experimentally. The in-medium width provides information on the strength of the $\eta^\prime N$ interaction, as studied in \cite{oset_ramos}, and it will be instructive to compare this result with in-medium widths obtained for other mesons. Furthermore, knowledge of the $\eta^\prime$ in-medium width is important for the feasibility of observing $\eta^\prime$ - nucleus bound systems theoretically predicted in some models \cite{Nagahiro}.
\section{Experiment and Data Analysis}
\label{sec:1}
\subsection{Experiment}
\label{sec:2}
The experiment has been performed at the ELSA facility in Bonn using the Crystal Barrel (CB) and TAPS detector system. The combined CB/TAPS detector covered 99\% of the full 4$\pi$ solid angle. The high granularity of the system makes it very well suited for the detection of multi-photon final states.  Tagged photon beams of energy up to 2.6 GeV were produced via bremsstrahlung and impinged on a solid target. For the measurements targets of  ${}^{12}\textrm{C}, {}^{40}\textrm{Ca}, {}^{93}\textrm{Nb}$ and ${}^{208}\textrm{Pb}$ were used. A more detailed description of the detector setup and the running conditions has been given in ~\cite{nanova,Elsner}.\\The detector acceptance was determined by Monte Carlo simulations using the GEANT3 package, including all features of the detector system, trigger conditions and all cuts for particle identification. To avoid further uncertainties due to reaction kinematics and final state interactions, the $\eta^\prime$-meson detection efficiency was simulated as a function  of the kinetic energy and the polar angle - $\epsilon(T_{\eta'}, \theta_{\eta'})$. Typical efficiencies are $\approx$ 7 \%, slighly different for the different targets. Experimental data are efficiency corrected event-by-event with this acceptance as described in ~\cite{igal,nanova2}.\\
\subsection{Reconstruction of the $\eta^\prime$-meson}
\label{sec:3}
 \begin{figure}[h!]
 \resizebox{0.9\columnwidth}{!}{
    \includegraphics{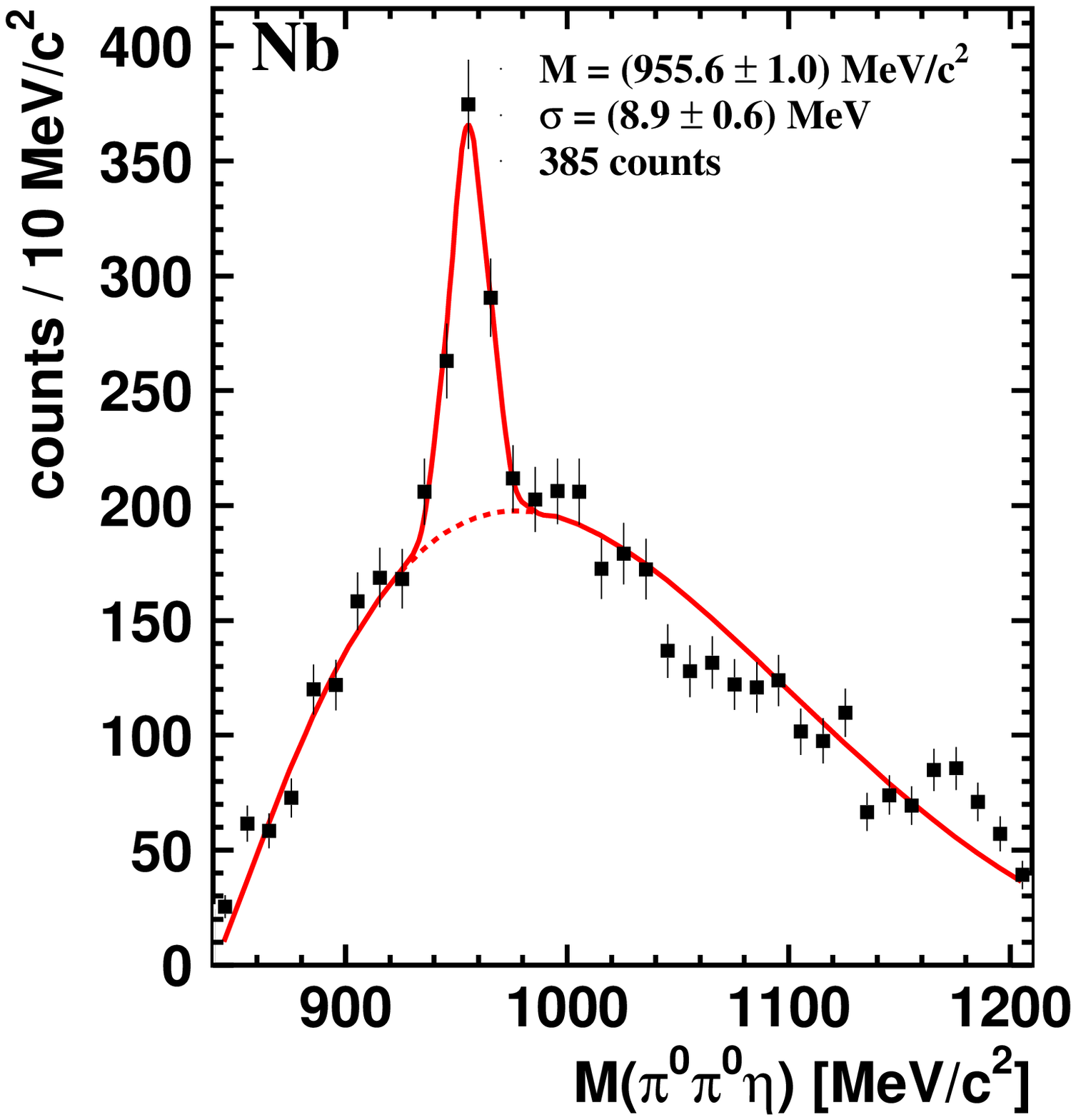} \includegraphics{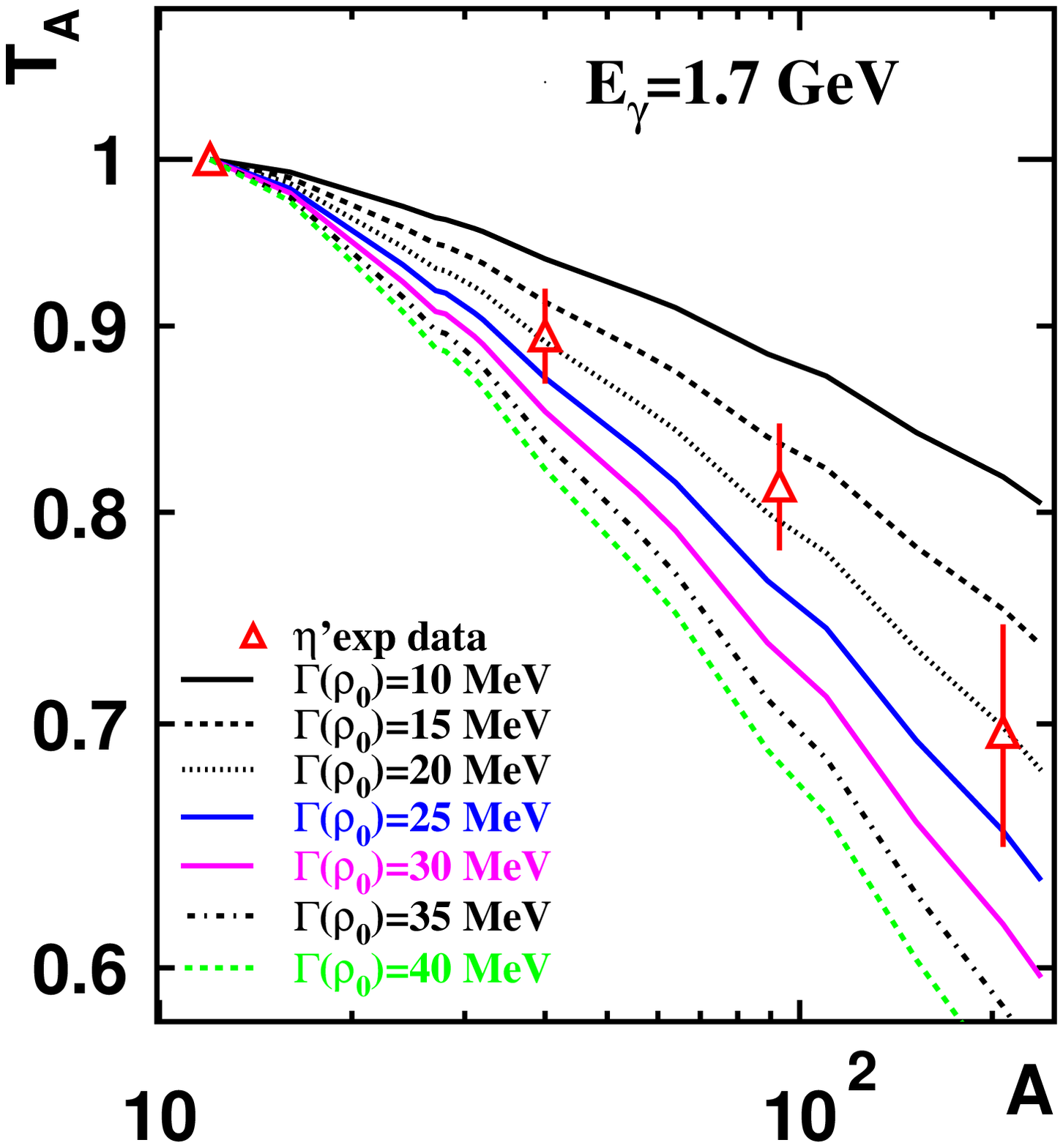} \includegraphics{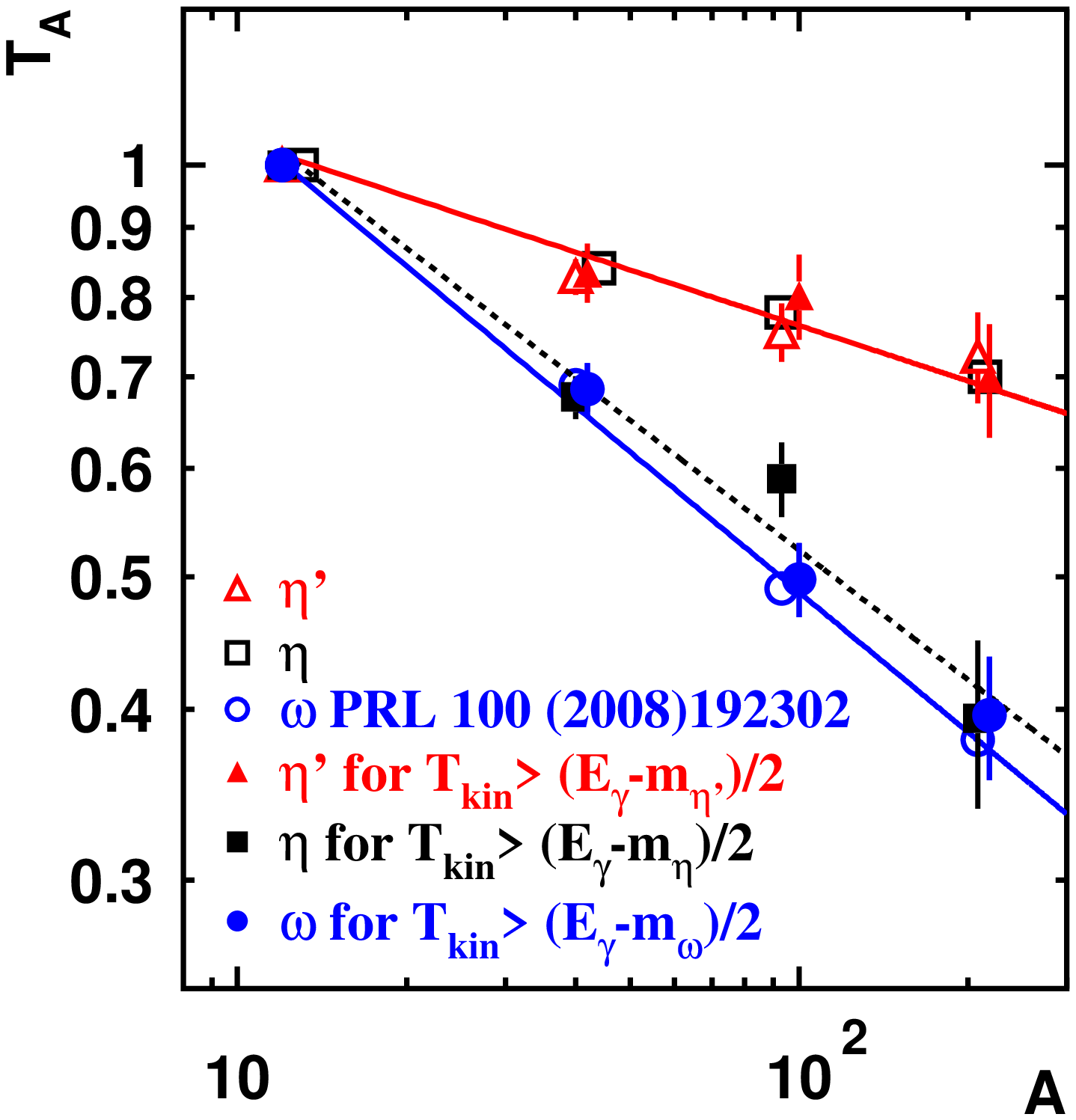}}
\caption{(left) Invariant mass spectrum of $\pi^{0}\pi^{0}\eta$ for the ${}^{93}\textrm{Nb}$ target for the incident photon energy range 1500 - 2200 MeV. The solid curve is a fit to the spectrum. Only statistical errors are given. (middle) Transparency ratio relative to that of $^{12}\textrm{C}$ (Eq.~\ref{eq:trans}), as a function of the nuclear mass number $A$ at incident photon energy range 1600-1800 MeV. The curves represent calculations for different in-medium widths of the $\eta^\prime$~\cite{nanova2}. (right) Transparency ratio for different mesons - $\eta$(squares), $\eta^\prime$(triangels) and $\omega$(circles) as a function of the nuclear mass number $A$.  The transparency ratio with a cut on the kinetic energy for the respective mesons is shown with full symbols. The incident photon energy is in the range 1500 to 2200~MeV. The solid and dotted lines are fits to the data.}
\label{fig:invmass}
\end{figure}
The $\eta^{'}$-mesons were identified via the $\eta^\prime \rightarrow \pi^{0} \pi^{0} \eta \rightarrow 6 \gamma$ decay channel, which has a branching ratio of 8\%. For the reconstruction of the $\eta^{'}$-mesons only events with at least 6 or 7 neutral hits have been selected. The competing channel with the same final state, namely $\eta \rightarrow \pi^{0} \pi^{0} \pi^{0} \rightarrow 6 \gamma$,  has been reconstructed and the corresponding events have been rejected from the further analysis. In addition only events were kept with at least one combination of the 6 photons to two photon pairs with invariant masses between 110 and 160 MeV ($\pi^{0}$) and one pair between 500 and 600 MeV ($\eta$). The $\pi^{0} \pi^{0} \eta$ invariant mass distributions for the different solid targets can be seen in ~\cite{nanova2}, here in Fig.~\ref{fig:invmass} (left) is shown only one example, namely the $\pi^0\pi^0\eta$ invariant mass measured on Nb. The resulting cross sections are used to calculate the transparency ratio of the $\eta^\prime$-meson for a given nucleus $A$.  
\subsection{Transparency ratio}
\label{sec:4}

Here, the production cross section per nucleon within a nucleus is compared to the production cross section of $\eta^\prime$ on a carbon target. $A_{eff}$ is hereby the effective number of participant nucleons reached by the photon beam which decreases due to photon shadowing relative to the total number of available nucleons with incident photon energy and target size. 
\begin{equation}
T^{C}_A=\frac{11 \cdot \sigma_{\gamma A\to \eta^\prime A^\prime}}{A_{eff}\cdot
\sigma_{\gamma C\to \eta^\prime X} } \ .
\label{eq:trans}
\end{equation} 
As shown in \cite{bianchi}, at the average photon energies of our experiment the shadowing of the photons results in an effective number of participant nucleons per nucleon of 0.88 for C and 0.84 for Pb. There is a difference of 5 $\%$ from C to Pb. This means that in the transparency ratio of Eq. (1), 5 $\%$ of the decrease of this ratio from C to Pb is due to the shadowing of the photons in the initial photon propagation and not due to the absorption of the $\eta^\prime$ in the final state interaction with the nucleus. To correct for this we increase the measured ratio by 5$\%$ for Pb and correspondingly  by 2 $\%$ for Nb and 1$\%$ for Ca, taking C as reference, which helps to suppress the distortion of the transparency ratio by photoabsorption and also by two-step processes~\cite{nanova2}.  
\section{Results and Discussion}
\label{sec:5}
\begin{figure}[h!]
 \resizebox{0.9\columnwidth}{!}{
    \includegraphics{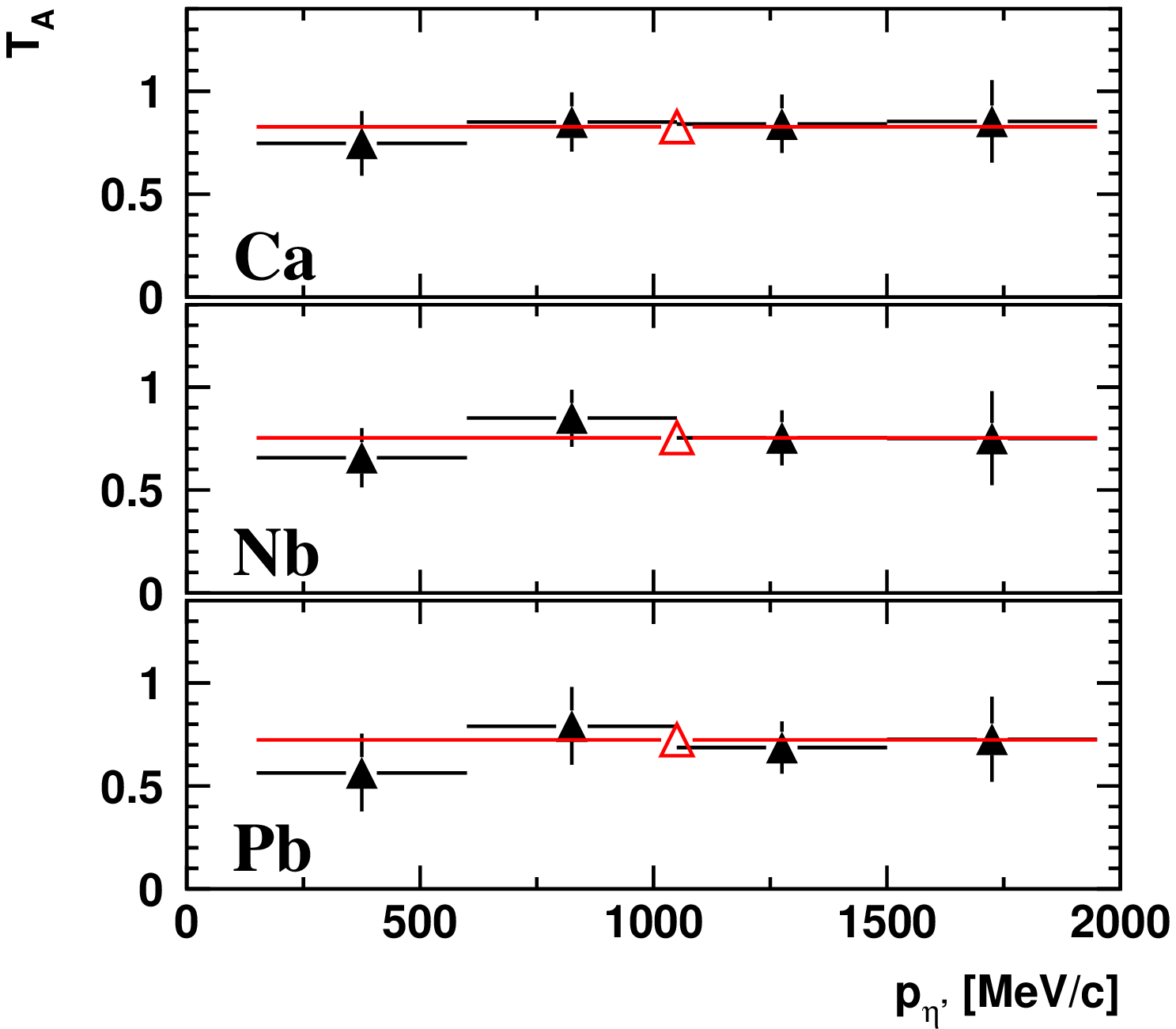} \includegraphics{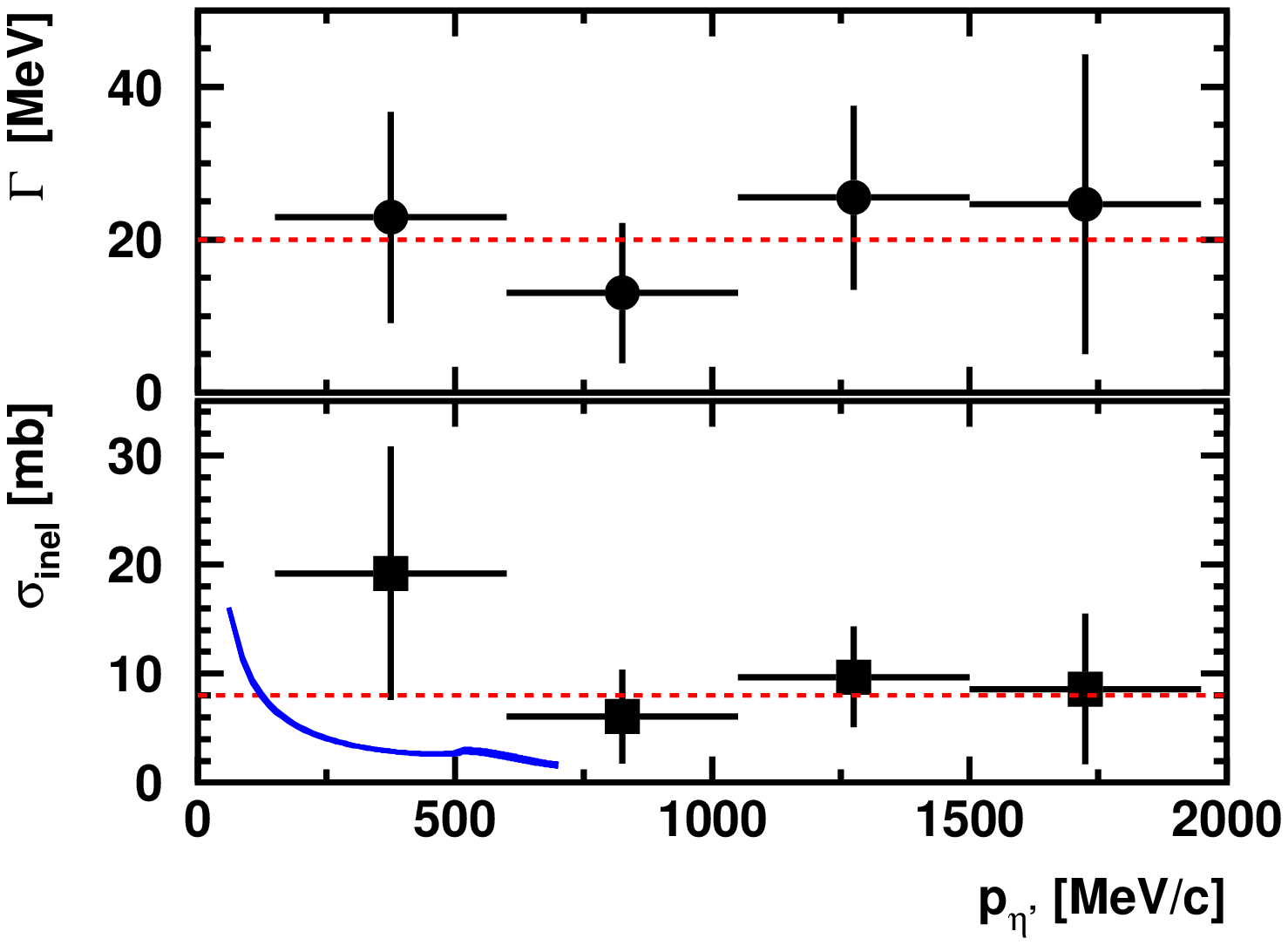}}
 \caption{(left) Transparency ratio for the $\eta^\prime$-meson normalized to C for three different targets: Ca, Nb, and Pb and four bins in $\eta^\prime$ momentum (full triangles) for the full incident photon energy range from 1500 - 2200~MeV. The open triangles show the values when integrated over all momenta and energies. (right) The in-medium width (upper panel) and inelastic cross section (lower panel) as a function of the $\eta^\prime$ momentum. The solid curve in the lower panel presents the calculation from ~\cite{oset_ramos}.}
 \label{fig:mom}
\end{figure}

The results are shown in Fig.~\ref{fig:invmass} (middle). The $\eta^\prime$ yield per nucleon within a nucleus decreases with increasing nuclear mass number and drops to about 70\% for Pb relative to carbon. A comparison of the data with calculations ~\cite{nanova2} 
indicates an in-medium width of the $\eta^\prime$ meson of 15-25 MeV at an average recoil momentum of $p_{\eta^\prime} $= 1.05\ GeV/c ~\cite{nanova2}. Assuming the low density approximation
\begin{equation}
\Gamma = \hbar c \cdot \rho_{0} \cdot \sigma_{\rm inel} \cdot \beta, \label{eq:Gamma-sigma}
\end{equation}
with
\begin{equation}
\beta = \frac{p_{\eta^\prime}}{E_{\eta^\prime}}
\end{equation}
in the laboratory and taking the average $\eta^\prime$ recoil momentum of 1.05 GeV/c into account, an inelastic cross section of $\sigma_{\rm inel} \approx$ 6-10~mb is deduced.
The contribution of secondary production processes like $\pi N \rightarrow  \eta^\prime N$ could increase the number of observed mesons and thus distort the transparency ratio measurement. The mesons reproduced via such processes should have relatively low kinetic energy. Applying a cut on the kinetic energy of the meson it is possible to study the effect of the secondary production processes on the transparency ratio. In Fig.~\ref{fig:invmass} (right) the results for the $\eta^\prime$-meson are compared to transparency ratio 
measurements for the $\eta$ \cite{thierry} and $\omega$ meson \cite{Kotulla}. The data are shown for the full kinetic energy range  of recoiling mesons (open symbols) as well as for the fraction of high energy mesons (full symbols) selected by the constraint
\begin{equation}
T_{kin} \ge (E_{\gamma} -m)/2. \label{cut}
\end{equation}
 Here, $E_{\gamma} $ is the incoming photon energy and $T_{kin} $ and $m$ are the kinetic energy and the mass of the meson, respectively. As discussed in \cite{thierry}, this cut suppresses meson production in secondary reactions. Fig.~\ref{fig:invmass} (right) shows that within errors this cut does not change the experimentally observed transparency ratios for the $\omega$-meson and $\eta^\prime$-meson while there is a significant difference for the $\eta$ meson.\\ 
The $\eta^\prime$ transparency ratio for different momentum bins has been determined and it is shown in Fig.~\ref{fig:mom} (left). The in-medium inelastic cross section and the in-medium width have been derived as a function of the $\eta^\prime$ momentum for each momentum bin separately. Within the errors no strong variation with momentum is observed as shown in Fig.~\ref{fig:mom} (right). This  indicates that two-step processes do not seem to play an important role in the photoproduction of $\eta^\prime$-mesons in the photon energy regime studied. This is an important observation because Eq.(\ref{eq:Gamma-sigma}) can only be applied to extract an inelastic cross section if two-step processes can be neglected. Otherwise the measured transparency ratio would reflect a convolution of secondary production and absorption in nuclei. In two-step processes where e.g. a pion is produced in the initial step by the incoming photon and the $\eta^\prime$-meson is then subsequently produced in a pion-induced reaction on another nucleon there is less energy available for the final state meson. This would shift the $\eta^\prime$ yield towards lower energies and lead to an enhancement of the transparency ratio at low $\eta^\prime$ momenta which is not observed.
 
\section{Conclusion}
\label{sec:6}
 From the transparency ratio measurement for the $\eta^\prime$-meson we find $\Gamma \approx  15-25$~MeV $\cdot \rho/\rho_0$ roughly, corresponding to an inelastic $\eta^\prime N$ cross section of $\sigma_{\rm inel} \approx$ 6-10~mb. Despite of the uncertainties and approximations involved in the determination of $\sigma_{\rm inel}$, this is the first experimental measurement of this cross section. A comparison to photoproduction cross sections and transparency ratios measured for other mesons ($\pi,\eta,\omega$) demonstrates the relatively weak interaction of the $\eta^\prime$-meson with nuclear matter. Regarding the observability of $\eta^\prime$ mesic states, the measured in-medium width of $\Gamma \approx  15-25$~MeV  at normal nuclear matter density would require a depth of about 50 MeV or more for the real part of the $\eta^\prime$ - nucleus optical potential.

\end{document}